\documentclass[12pt,preprint]{aastex}
\slugcomment{Accepted and scheduled for publication  
in {\it The Astrophysical Journal},  for the ApJ July 1, 2007, v 663, 1 issue}  
\def\lax {\ifmmode{_<\atop^{\sim}}\else{${_<\atop^{\sim}}$}\fi}  
\def\gax {\ifmmode{_>\atop^{\sim}}\else{${_>\atop^{\sim}}$}\fi}  
\def\gtorder{\mathrel{\raise.3ex\hbox{$>$}\mkern-14mu
             \lower0.6ex\hbox{$\sim$}}}
\def\etal { et al. }

\begin{document}

\title{Determination of Black Hole Mass in Cyg X-1 by Scaling of
Spectral Index-QPO Frequency Correlation }

\author{Nickolai Shaposhnikov\altaffilmark{1,2} and Lev Titarchuk\altaffilmark{2,3} }

\altaffiltext{1}{CRESST/Universities Space Research Association, 10211 Wincopin Cir, Suite 500, Columbia MD, 21044, nikolai@milkyway.gsfc.nasa.gov}

\altaffiltext{2}{Goddard Space Flight Center, NASA, 
Astrophysics Science Division, code 662, Greenbelt MD 20771}

\altaffiltext{3}{George Mason University/Center for Earth
Observing and Space Research, Fairfax, VA 22030; and US Naval Research
Laboratory, Code 7655, Washington, DC 20375-5352; lev.titarchuk@nrl.navy.mil;  
Goddard Space Flight Center, NASA,  code 661, Greenbelt  
MD 20771; lev@milkyway.gsfc.nasa.gov}

\begin{abstract}
It is well established  that timing and spectral properties of
Galactic Black Hole (BH) X-ray binaries (XRB) are strongly correlated.
In particular, it has been shown that low frequency 
 Quasi-Periodic Oscillation (QPO)  $\log \nu_{low}$ -
photon index  $\Gamma$ correlation curves have a specific pattern. In a number of  sources the shape of the index-low frequency QPO correlations  are self-similar with a position offset  in the  $\log\nu_{low}-\Gamma$ plane. Titarchuk \& Fiorito  presented strong  theoretical and observational  arguments that the QPO frequency values  in this $\log\nu_{low}-\Gamma$ correlation should be  inversely proportional to $M_{BH}$.  A simple translation 
of the correlation for a given source  along the frequency axis leads to the observed correlation for another source. As a result of this translation one can  obtain  a scaling factor which is simply a  BH mass ratio for these particular sources. 
This  property of the correlations offers
a fundamentally new method for BH mass determination in XRBs. 
Here we use the observed QPO-index correlations
observed in three BH sources: GRO J1655-40, GRS 1915+105 and Cyg X-1.
The BH mass of $(6.3\pm 0.5)$ M$_{\odot}$ in   GRO J1655-40 is obtained using  optical observations.
 {\it RXTE} observations during the
recent 2005 outburst yielded sufficient data to establish the correlation
pattern during both rise and decay of the event.
We use GRO J1655-40 as a standard reference source to measure the BH mass 
in Cyg X-1. We also revisit the GRS 1915+105 data as a further test of our scaling method. We infer
the value of BH mass of $(15.6\pm 1.5)M_{\odot}$ in this source which  is consistent with the previous BH mass estimate in GRS 1915 of $(13.3\pm 4)M_{\odot}$.
We obtain the BH  mass in Cyg X-1 in the range $(8.7\pm 0.8)M_{\odot}$. 

\end{abstract}


\section{Introduction}
In Galactic Black Hole (GBH) sources transitions between spectral states are accompanied by
a characteristic evolution of timing properties \citep[see][for the review of BH spectral states]{mr}.
In particular, low frequency quasi-periodic oscillations (LF QPO) observed during low-hard
and intermediate states are very closely correlated with the photon index of the power law 
spectral component \citep[e.g.][]{vig}. Recently, the same type of correlation have been firmly
established for Cyg X-1 [\citet{ST06}, hereafter ST06] and GRO J1655-40 [\citet{sh07}, hereafter S07]. In this Paper  we 
use these correlations to constrain the mass of BH in Cygnus X-1 based on the theoretically
motivated dependence of QPO-index pattern on the mass of the central BH. 

Some progress has been made in understanding how variability scales with BH mass
between GBHs and AGNs \citep[see][and references therein]{dg03}.
However, in most studies authors concentrate on
the observational aspects giving less attention to the physical reasoning 
behind the scaling method. The ubiquitous nature of the correlation of index with QPO frequency 
 suggests that the underlying physical process which gives rise 
to the LF QPO is closely tied to the corona; and, furthermore, 
that this process varies in a well defined manner as a source progresses 
from one spectral  state to another. 
Moreover, the fact that the same correlations are seen in so many sources, 
which vary widely in both luminosity (presumably with mass accretion rate) and state, suggests that 
the physical conditions controlling the index and the low frequency QPOs are characteristics of these sources, and that by 
virtue of the low-high-frequency correlations [see Psaltis, Belloni \& van der Klis (1999)   
 and  Belloni, Psaltis \& van der Klis 2002), they  may be a universal property 
of  all  accreting compact systems.
The observational evidence presented above motivated the development of a detailed 
physical model of the corona surrounding a BH which directly predicts the behavior of the
spectral index with fundamental properties of the corona [see Titarchuk, Lapidus \& 
Muslimov (1998), hereafter TLM98; \citet{lt99}; Titarchuk \& Osherovich (1999) and 
Titarchuk  \& Fiorito (2004), hereafter TF04).  This  model incorporates the fundamental principles 
of fluid mechanics, radiative
transfer theory and oscillatory processes. It identifies the LF QPOs
as normal mode oscillation frequency of a compact coronal region near the BH and shows
how the photon index of this corona changes as a function of mass accretion rate.
It is important to emphasize  that the variable QPO frequency scaling as $1/M$ is a generic feature
of  the TLM98 model.  In fact, the  QPO frequency scales as a ratio of plasma (magneto-acoustic) 
velocity and  {\it  the  size of the oscillating region, which is measured in the Schwarzchild radius units.}

TF04  illustrate how the mass of one BH source can be used to determine
that of another using the {\it RXTE} data from GRS 1915+105 and XTE J1550-564,
which exhibit remarkably similar QPO frequency$-$index correlation curves.  
The scaling factor between the LF QPOs of two X-ray BH sources  with
similar power-law (PL) photon indices can be used to determine the ratio
of their BH masses. The mass determination using the QPO
frequency$-$index correlation  is consistent with X-ray spectroscopic
and dynamical mass determinations for these sources 
(see  references in Shrader \& Titarchuk 2003).
 The scaling was successfully applied
for ultra-luminous sources using QPO frequencies \citep{ft04,str06}.
\citet{ft04} and \citet{dtg06}  applied  this  new method of BH mass 
determination  to estimate the black hole mass of the ultraluminous X-ray sources M82 X-1. 
Using scaling arguments and the correlation derived from the consideration of Galactic black holes, they  
conclude that M82 X-1 is an intermediate black hole with a mass in the range of (250-1000) M$_\odot$. 
Recently this technique led \citet{str06} to estimation of BH mass of $\sim 1500\, M_{\odot}$ in ULX source NGC 5408 X-1. 

In this Paper we demonstrate how the  TF04 scaling method can be applied to the determination of the 
Cyg X-1 BH mass $M_{Cyg X-1}$.   We obtain  $M_{Cyg X-1}$ when we scale the index-QPO correlation  of  GRO J1655-40 vs that of Cyg X-1 along frequency axis.  We test this scaling method using BH masses of  GRS 1915+105 and GRO J1655-40 previously inferred  from optical and IR observations. 

\section{Sources and Observations}

\subsection{GRO J1655-40}
The X-ray binary 
GRO~J1655--40 (GRO J1655 hereafter) is a well-known example of a BH X-ray transient that 
underwent several major outbursts within  the last 12 years.
It was discovered by the BATSE instrument onboard the 
Compton Gamma-Ray Observatory in mid 1994 \citep{zh94}.
The secondary star being relatively bright, the
binary parameters are exceptionally well 
determined among LMXBs. 
The most recent optical  photometry of \citet{greene01} led to 
 a BH mass estimate of $(6.3\pm0.5) M_\odot$. 
During {\it RXTE} era GRO J1655 had two major outburst in 1996/97 and in 2005. 
However, during the first event pointed {\it RXTE} observations started
when the source was already in the high-soft  state  and the state transition was
not covered. The 2005 outburst was recognized at a very early stage and 
observers at different wavelengths were relatively prepared and successful at 
responding relatively quickly (see S07, for extended report on multiwavelength 
campaign during 2005 outburst). Almost daily {\it RXTE} coverage allowed to
observe  timing and spectral evolution of the source in detail. We distinguish the data
during the outburst rise and decay because the correlation patterns are slightly offset
presumably due to different plasma temperatures in the corona (see below for explanation).
The data during the outburst rise were collected between MJD 53420 and 53440, while
the decay data were collected between MJD 53627 and 53640.  We refer a reader to  S07 for the 
details of spectral and timing data analysis of the outburst rise data. 
The decay data were analyzed in the same manner.    

\subsection{Cyg X-1}

Cyg X-1 is one of the brightest high-energy sources in the sky, with 
an average 1-200 keV energy flux of $\sim3\times10^{-8}$ ergs
cm$^{-2}$s$^{-1}$. Its optical companion is an O9.7 Iab supergiant HDE 226868.
 Estimates of the mass of the X-ray star, $5\lax M_{\odot}\lax15$
[e.g., \citet{her95}] strongly suggest the presence of a black hole. 
Observed spectral and temporal X-ray characteristics are
extensively studied by ST06,  based on the large amount of  data  collected in 
the {\it RXTE} archive (see  \S 2 in ST06  for the data
description).  They  presented  the  analysis which includes $\sim$ 2.2 Ms of {\it RXTE} archival data from Cygnus X-1. 
The main findings of the ST06  study are following:
The photon index $\Gamma$ steadily increases from 1.5
in the low/hard state to values exceeding 2.1 in the high-soft state. 
The low frequency $\nu_L$ is detected throughout the low-hard and intermediate state, while it disappears when the source
undergoes transition to the soft (thermal dominated) state.
Like in other BH sources, there is  an indication of saturation
of the index in the high-soft state for Cyg X-1 (see ST06, Fig. 8) 
This saturation 
effect, which  is presumably due to photon trapping in the converging flow, 
can be  considered to be a BH signature.

\subsection{GRS 1915+105}

GRS 1915+105 (GRS 1915 hereafter) is another prominent Galactic BH binary. The source is very bright
and exhibits  very diverse temporal properties. In attempt to classify the variability in GRS 1915 \citet{bel00}
identified at least 12 different classes. \citet{mrg97} reported  
 three different types of QPO: a QPO with constant centroid frequency of 67 Hz, a dynamic 
low-frequency ($10^{-3}-10$ Hz) QPO with a large variety of amplitudes and widths, and high-amplitude
``sputters'' at frequencies of $10^{-3}-10^{-1}$ Hz. Authors also mention that it is difficult to 
describe the source properties in term of conventional BH states. A BH mass estimation for GRS 1915+105
of $13.3\pm 4$ is available from IR observations [\citet{gre01}] and X-ray observations [\citet{BRT99} and \citet{shT03}].

Despite the complicated nature of the  GRS 1915 spectral and timing properties we were able to
construct the QPO-photon index correlations. We use the data from six {\it RXTE} observations, namely:
10408-01-27-00, 20402-01-50-00, 20402-01-50-01, 20402-01-51-00, 20402-01-52-000, 40702-01-02-00.
All the data, except for the last observation, are collected in the hard state. During this
state  power density spectrum (PDS) of the source is very close in shape to the PDS of the 
classical hard state ($\chi$ Belloni's variability class),
 i.e. the band-limited noise plus one or two low frequency QPOs. 
Data from these observations yield the lower part of QPO-index correlation curve (see Fig. \ref{1915_1655_shift}). 
During the observation 40702-01-02-00 the source exhibited irregular variability with flaring episodes ($\nu$ class). This observation yields the upper part of the correlations with a saturation. 
For each observation we calculated energy and Fourier spectra for consecutive intervals of 128 seconds.  Following the approach of \citet{vig} we fit the data with
a sum of multicolor disk and power law models. We fit PDS continuum with broken power law shape to model band-limited noise and we modeled QPOs with Lorentzians. 

\section{QPO-spectral index correlation model}

In most  BH sources, the correlation of the spectral index$-$QPO frequency  follows a remarkably similar pattern. Namely, at lower QPO  frequencies  the dependence starts at the index of  $\sim 1.5$ (e.g. low-hard state) and follows linear correlation with positive slope until some QPO frequency value $\nu_{br}$ where it levels off. This saturation of  index versus QPO frequency 
approximately corresponds  to the  intermediate state.  Then, upon the QPO turn off,
the source enters the soft state. 


According to our scenario (see TLM98 and TF04 for details) the LF QPO features  are  observational appearances  of  a Compton cloud  (CC) undergoing 
normal mode volume oscillations.   The radius of the  CC outer boundary is the  radius where adjustment   
of disk Keplerian rotation to sub-Keplerian CC rotation occurs. It  should be  determined  in  Schwarzchild radius ($R_{\rm S}$)  units, namely $R_{out}=r_{out} R_{\rm S}$. Thus the CC size
should be  proportional to $m=M/M_{\odot}$ 
  and  QPO frequency (as a ratio of the magneto-acoustic velocity and size) should  be  reciprocal to 
the characteristic size of the system and consequently to $m$.  The spectral index, on the other hand,
is a measure of CC upscattering  efficiency.  

Note that direct connection of  variability in the low-hard and intermediate states with 
power law spectral component  has been identified and well established in many studies \citep[see][and references therein]{klis95}.   
In the hard state the mass accretion rate is low, the size of the corona is large and the
index is determined by thermal Comptonization in an optically thin geometrically thick
configuration. When accretion rate starts to increase a cold disk becomes stronger
and CC cooling by disk photons becomes more efficient. Ultimately the CC plasma temperature goes down.  In the high-soft state the CC temperature  is of the order of the disk temperature (about 1 keV).   As the mass accretion rate increases the adjustment radius decreases and consequently the CC cloud contracts  (see TLM98, TO99).  QPO frequency increases with the CC contraction,  while decreasing
CC temperature leads to the lower  efficiency of Comptonization (lower Comptonization parameter $Y$),  and thus to higher  photon indices $\Gamma$. Bradshaw, Titarchuk \& Kuznetsov (2007) derived a  simple relation which shows that  $\Gamma$ anti-correlates with $Y$,  namely they show that $\alpha=\Gamma-1=Y^{-1}$.

The presented picture  explains a positive correlation between spectral index and QPO frequency.
The index saturation for higher QPO values can be accounted for by the bulk motion inflow (dynamical)  
Comptonization which becomes dominant for the final stage of state transition when 
the corona is  cooled down by the disk photons.   The saturation level of index can vary 
for the same source for different transition episodes. This index saturation  value is determined 
 by the plasma temperature of  the converging flow (CF) [see \citet{lt99}]
during a  transition.  On the other hand  (see above) the CF plasma temperature strongly depends on the CF illumination by the disk photon. For  higher photon  flux from the disk,  the CF plasma  temperature is lower.  The level of the index  saturation decreases when  the CF plasma temperature increases. 

Thus the index  saturation levels can be  different from source
 to source depending on the strength of the disk in these sources. 
 Even for the same source the index saturation  can be variable. In Figures  \ref{1915_1655_shift} and   \ref{cygx1_1655_shift}  for GRO J1655-40  one can see the different index saturation levels 
 $\Gamma_{sat}\sim 2.3$, and $2.05$  during 2005 outburst rise and decay stages respectively. 
More details of this physical scenario theory  can be found in  \citet{tF} (and references therein). 

We must stress  that  the main property of index-QPO relation which we use for scaling method
{\it does not depend on a specific physical mechanism of photon  and  power density spectral  formation}.
The only assumption we use is that for a given index  QPO frequency is inversely proportional  to a 
 BH mass (see TLM98, TF04).   
Important advantage of the technique is its independence of
the system geometry and a distance to a source.

The disadvantage of the method is that the mass of one BH exhibiting clear index-QPO pattern
has to be well known. Fortunately, such a system is available, i.e. GRO J1655-40. 
Another restriction of the method is that it is not applicable to the data sets with
different index saturation  levels.  The main requirement of the scaling
method is the QPO-index  correlations should be self-similar for two  given sources.  In the sense that  the  scaling one correlation versus another along log-frequency axis leads to the  same shape for these sources. The scaling method  does not work properly  for the data sets with  different index  saturation levels.

\section{Data analysis. Application of scaling method}

In principle, it is possible to construct index-QPO relation 
on the basis of the physical  theory presented in TF04. However,
 the exact relation between optical depth $\tau$
and Reynolds number $\gamma$ (which is the inverse of $\alpha$-parameter in the disk)
 requires the knowledge of the disk equation of state. 
 The direct implementation of
the theory has to  include some assumption about the functional form of $\tau(\gamma)$ because of uncertainties of the disk equation of state and the plasma viscosity  in the disk.  
It is worth noting  that any $\alpha-$disk model does not specify an exact value of $\alpha-$ parameter 
[see e.g. \citet{ss73}; \citet{nar94}; \citet{CT95}].

The fact that the scaling method  follows from the first principles 
allows us to avoid the full implementation of the accretion theory to construct QPO-index relation for each particular source. 
All we need is a simple function which correctly represens the data and gives us means to calculate a scaling coefficient between two correlation curves.
The correlations between QPO frequency and photon index for three sources considered in 
this Paper are shown in Figures \ref{1915_1655_shift} and  \ref{cygx1_1655_shift}. 
It is clear that QPO-index correlations follow a simple pattern. Namely, the dependence is 
linear with particular slope for lower frequencies, while near some frequency (which we call
$\nu_{tr}$) the function smoothly levels off to become a constant.
This behavior is well reproduced by the following  analytical function:
\begin{equation}
f(\nu)=A-D\,B\ln\{ \exp [(\nu_{tr}-\nu)/D]+1\}.
\label{anat_corr}
\end{equation}
The two asymptotes of this function are $f(\nu)=A+B(\nu-\nu_{tr})$ for $\nu<\nu_{tr}$ and
$f(\nu)=A$ for $\nu>\nu_{tr}$. Thus, 
$A$ is a value of the index saturation level, $B$ is a slope of the low frequency part of the data, 
$\nu_{tr}$ is the  frequency at which  index-QPO dependence levels off . Parameter $D$ 
controls how fast the transition occurs. 
It is important to point out  that the slope $B$ of the asymptote  for $\nu<\nu_{tr}$ in Eq. (\ref{anat_corr}) is proportional to the BH mass.
Using the above model function we fit the observed index-QPO correlations
and apply the scaling method for mass BH determination.

\subsection{Scaling from  GRO J1655-40 correlation to  GRS 1915+105 correlation}

\begin{figure}[ptbptbptb]
\includegraphics[scale=0.6, angle=-90]{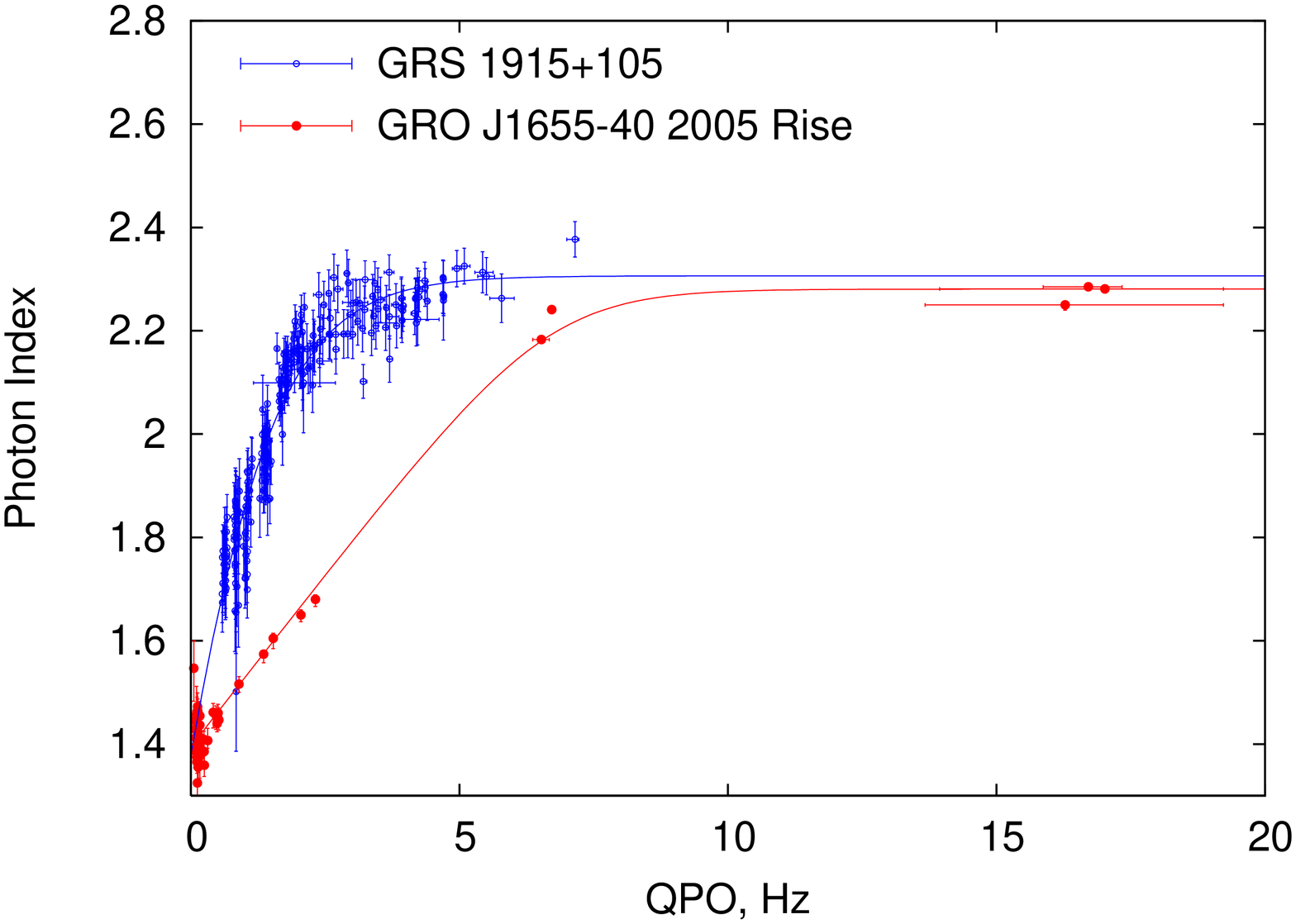}
\caption{Spectral Index-QPO frequency correlations data for GRS 1915+105 (blue points) and GRO J1655-40 
during the 2005 outburst rise (red points).  A clear saturation of index for high values of QPO is seen 
for both sources. The lower part of correlation is not available for GRS 1915+105. The data are fitted 
with with a model function to apply a scaling method. We obtain the BH mass in GRS 1915+105 in the range  of $(15.6\pm 1.5)\,M_\odot$  using  the scaling method  and the well-known BH mass of GRO 1655-40 which is in the range of   $6.3\pm0.5\,M_\odot$. 
}
\label{1915_1655_shift}
\end{figure}

\begin{figure}[ptbptbptb]
\includegraphics[scale=0.6, angle=-90]{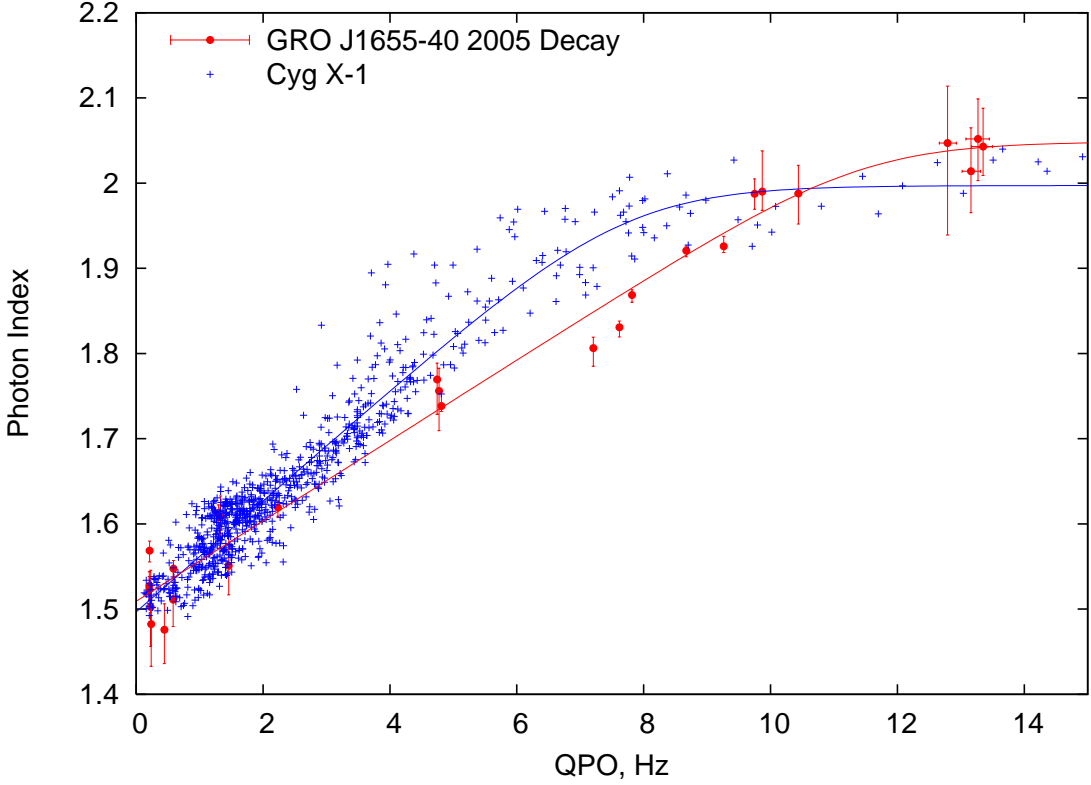}
\caption{Determination of BH mass in Cyg X-1. Correlations for Cyg X-1 
comprises {\it RXTE}-mission-long data from ST06 (blue color). Data from GRO 1655-40 is for the 2005 outburst decay. These sources show similar high frequency saturation levels for each of the correlations. Index-QPO correlation scaling gives us the 
BH mass in Cyg X-1 in the range of $(8.7\pm0.8)\,M_\odot$.
}
\label{cygx1_1655_shift}
\end{figure}

For scaling with GRS 1915 we choose the data from the rise
of the 2005 GRO J1655 outburst because in both cases the photon index saturation 
level is approximately  $\Gamma\approx 2.3$. We first fit the GRO J1655 points. We   found the following best fit parameters:
$A_{J1655}=2.28\pm0.02$, $B_{J1655}=0.133\pm 0.008$ Hz$^{-1}$ and $\nu_{tr,J1655}=6.64\pm0.48$ Hz. The $D$ was fixed at 1.0 Hz.
The best-fit curve is shown by a red solid line in Figure \ref{1915_1655_shift}. 
The best fit values  
found
  for GRS 1915+105 are: $A_{1915}=2.25\pm0.01$, $B_{1915}=0.33\pm 0.02$ Hz$^{-1}$ and  $\nu_{tr,1915}=2.23\pm0.07$ Hz.
For a scaling factor we have $s_{J1655\rightarrow 1915}=B_{1915}/B_{J1655}=2.5\pm0.21$. 
We obtain  a BH mass in GRS 1915 as 
\[
M_{1915}=B_{1915}\times M_{J1655}/B_{J1655}=(15.6\pm 1.5)M_{\odot},
\]
using the scaling method and  the value of BH mass in GRO  J1655 of $(6.3\pm0.5)M_{\odot}$.
The inferred uncertainties (error bars) of $M_{1915}$ are mostly affected by scattering and error bars of 
  index-QPO points in the GRS 1915 correlation (see Fig. \ref{1915_1655_shift}). 
The value of BH mass of $(15.6\pm 1.5)M_{\odot}$  is consistent  with the 
 previous BH mass estimate in GRS 1915 of  $(13.3\pm4)M_{\odot}$.   
This result of QPO-index correlation  scaling 
demonstrates a potential power  of our method of a BH mass determination.

\begin{figure}[ptbptbptb]
\includegraphics[scale=0.6,angle=-90]{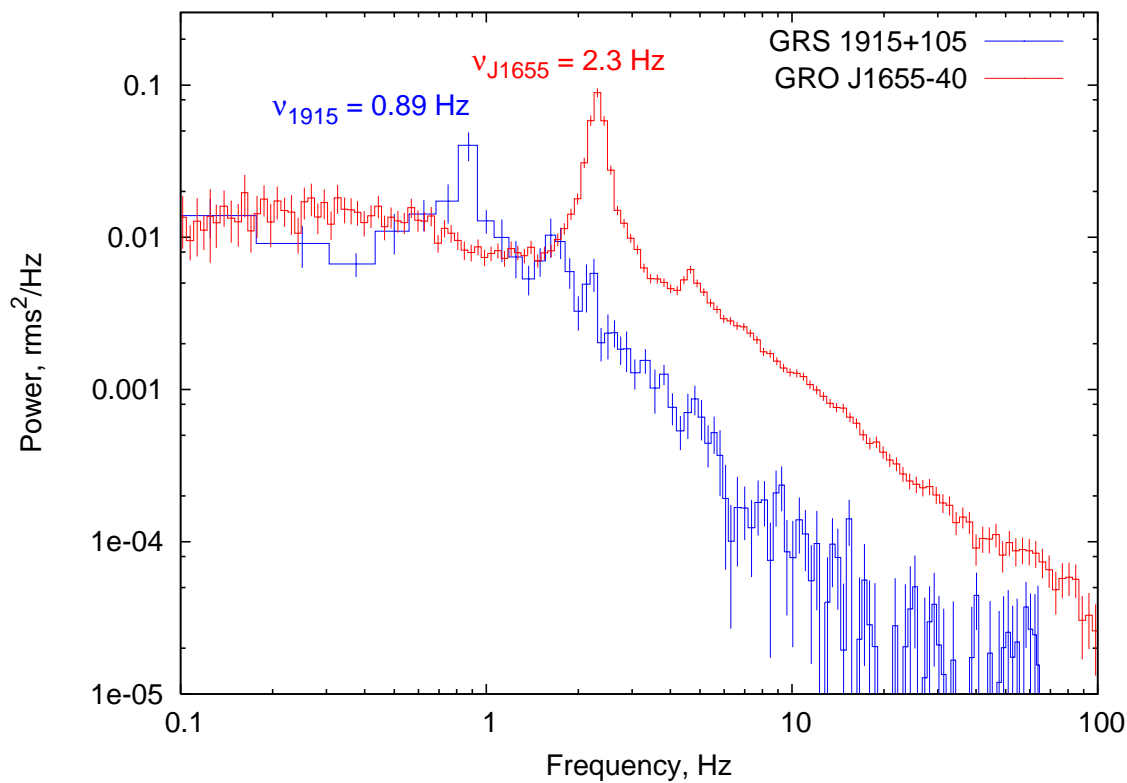}
\caption{ Individual PDS for GRS 1915+105 ({\it RXTE} Observation ID 20402-01-50-00, 128 sec. interval, see text for details) 
and GRO J1655-40 ({\it RXTE} Observation ID 90704-04-01-00) illustrating how low-frequency QPOs 
scale with the mass of the central black hole. The photon indices are $1.68\pm0.02$ and $1.67\pm0.08$ correspondingly.
PDS have the same flat-top noise levels and similar slopes after break but QPOs for GRO 1655-40 are shifted by approximately 
a factor of $ 2.5$.
}
\label{1915_1655_pds}
\end{figure}

In Figure \ref{1915_1655_pds} we illustrate the idea of the scaling method using  individual PDSs related to the same spectral state (index).
The PDSs presented for GRO J1655 and GRS 1915  have the same flat-top noise levels and similar slopes after the break. However, QPOs for GRO 1655-40 are shifted by approximately 
a factor of $\sim 2.5$.

\section{Scaling from GRO 1655 to Cyg X-1 correlations.  BH Mass  in Cyg X-1.}

Now we proceed with the main result of this  Paper, namely,  with the mass determination of BH 
 in Cyg X-1 using QPO-index scaling. We apply exactly the same procedure
(described in the previous section)  for the {\it RXTE} mission-long data of Cyg X-1 and 
the GRO J1655 data from decay stage of its 2005 outburst. We perform the initial fit to  Cyg X-1 
data to obtain: $A_{Cyg X-1}=2.00\pm0.01$, $B_{Cyg X-1}=0.065\pm 0.001$ Hz$^{-1}$ and $\nu_{tr,Cyg X-1}=7.68\pm0.15$ Hz.
Fit for GRO J1655 decay data gives: $A_{J1655}=2.05\pm0.02$ and $B_{J1655}=0.047\pm 0.002$ Hz$^{-1}$ and $\nu_{tr,J1655}=11.4\pm0.6$ Hz. For both fits we freeze $D$ at 1.0.
For the mass   of BH in Cyg X-1 we have:
\[
M_{Cyg X-1}=B_{Cyg X-1}\times M_{J1655}/B_{J1655}=(8.7\pm 0.8)M_{\odot} 
\]
Here we again  use $(6.3\pm0.5)M_{\odot}$ as a value of BH mass in GRO J1655.  
This inferred   value of Cyg X-1 BH mass  is slightly lower than 10 $M_\odot$ given by \citet{her95} (with no error bars quoted) and it is within the  mass range found by \citet{gb86}. 

\begin{figure}[ptbptbptb]
\includegraphics[scale=0.6,angle=-90]{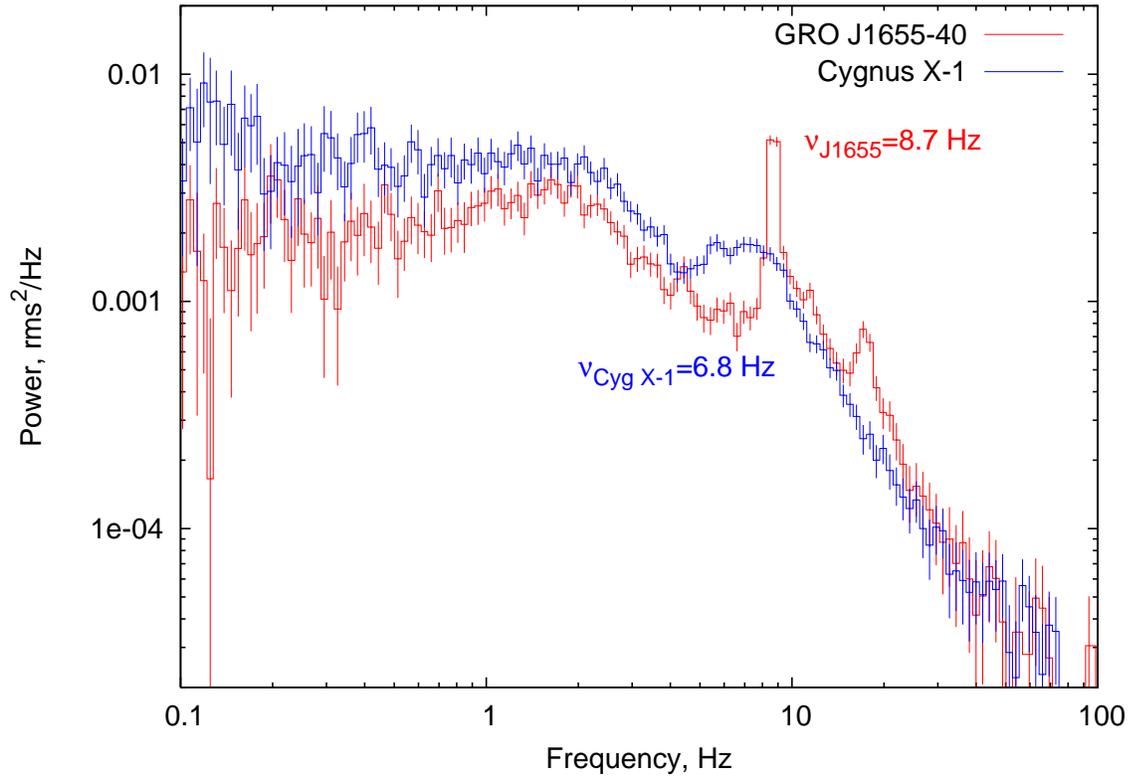}
\caption{Individual PDSs for Cyg X-1 ({\it RXTE} Observation ID 50119-01-03-01) and GRO 1655-40 ({\it RXTE} Observation ID 91702-01-79-00).
Spectral index for both observations is  about 1.92. 
QPO  feature in Cyg X-1 is less coherent than that for GRO 1655-40 which is presumably due to scattering in strong ambient
wind/corona (see ST06 for details).} 
\label{cygx1_1655_pds}
\end{figure}

In Figure \ref{cygx1_1655_pds} we present the individual PDS of Cyg X-1 and GRO J1655 related 
to the same spectral state (index). One can see that the corresponding  values of QPO frequencies are very close. However QPO in Cyg X-1 is less coherent than that in GRO 1655 which is presumably a result of photon scattering in the strong ambient  wind of Cyg X-1 (see ST06 for details). 

\section{Conclusions}
Our  results for the QPO-index correlation  scaling 
from  the GRO J1655 to  GRS 1915 and from the GRO J1655 to  Cyg X-1 data points
demonstrates a potential power  of the new method of a BH mass determination.

We have tested this scaling method using the known (from optical and IR observations) BH masses in  GRO J1655 and   GRS 1915. We were able 
to reproduce and confirm the previous estimate of BH mass of GRS 1915 $M_{1915}$. We obtain that $M_{1915}/M_{\odot}=15.6\pm1.5$. 

Our scaling method along with the observed index-QPO correlation give for the BH mass in  Cyg X-1  $M_{Cyg X-1}=(8.7\pm 0.8)M_{\odot} $, which is the most close constrains on the BH mass for this 
prominent source  obtained  up to date in  literature. Low Cyg X-1 mass function is an obstacle for dynamical mass measurement from optical data, which is also affected by the uncertainty  in the source distance. Our alternative method does not suffer from these restrictions, which allows such a precise measurement of the BH mass. 

We acknowledge the referee's constructive suggestions for the improvement of the paper presentation.

\newpage

\newpage

\newpage

\end{document}